\documentclass[reqno]{amsart}
\usepackage{array}
\numberwithin{equation}{section}  
 \DeclareMathOperator{\diag}{\rm diag}
 \DeclareMathOperator{\ad}{\rm ad}

 \DeclareMathOperator{\Ss}{\mathbb{S}}
 \DeclareMathOperator{\Lo}{\mathcal{L}}
 \DeclareMathOperator{\q}{\mathbf{q}}
 
\begin{document}
\title[]{Commutator
identities on associative algebras and integrability of nonlinear pde's}
\author[]{A.~K.~Pogrebkov}
\address[A.~K.~Pogrebkov]{Steklov Mathematical Institute, Moscow, Russia}
\email{pogreb@mi.ras.ru}
\thanks{This work is supported in part by the Russian Foundation for Basic Research (grant \# 05-01-00498),
Scientific Schools 672.2006.1, by NWO--RFBR (grant \# 05-01-89006), by the Program of RAS ``Mathematical
Methods of the Nonlinear Dynamics.''}
\date{\today}

\begin{abstract}
It is shown that commutator identities on associative algebras generate solutions of linearized integrable
equations. Next, a special kind of the dressing procedure is suggested that in a special class of integral
operators enables to associate to such commutator identity both nonlinear equation and its Lax pair. Thus
problem of construction of new integrable pde's reduces to construction of commutator identities on
associative algebras.
\end{abstract}

\maketitle

\section{Introduction}

In~\cite{pogr} we mentioned that any two arbitrary elements, $A$ and $B$, of an arbitrary associative
algebra obey the following commutator identity:
\begin{equation}
 [A_{}^{3},[A,B]]-\dfrac{3}{4}[A_{}^{2},[A_{}^{2},B]]-\dfrac{1}{4}[A,[A,[A,[A,B]]]]=0.\label{commut1}
\end{equation}
Defining adjoint action of powers of element $A$ on the associative algebra as
\begin{align}\label{ad1}
  &\ad_{n}B=[A^{n},B] \\
\intertext{one can write~(\ref{commut1}) as relation between these adjoint actions:}
 &\ad_{3}\ad_{1}-\dfrac{3}{4}(\ad_{2})^{2}-\dfrac{1}{4}(\ad_{1})^{4}=0.\label{ad2}
\end{align}
Being the trivial consequence of associativity identity~(\ref{commut1}) readily proves that function
\begin{equation}\label{Bt1}
  B(t_1,t_2,t_3)=e^{t_1A+t_2A^2+t_3A^3}_{}Be^{-t_1A-t_2A^2-t_3A^3}_{}
\end{equation}
obeys linearized Kadomtsev--Petviashvili (KP) equation~\cite{KP} with respect to variables $t_j$, i.e.,
\begin{equation}
 \dfrac{\partial_{}^{2}B(t)}{\partial t_{1}^{}\partial t_{3}^{}}-
 \dfrac{3}{4}\dfrac{\partial_{}^{2}B(t)}{\partial t_{2}^{2}}-
 \dfrac{1}{4}\dfrac{\partial_{}^{4}B(t)}{\partial t_{1}^{4}}=0,  \label{KP}
\end{equation}
or more exactly, KPII. In~\cite{pogr} it was pointed out that there exist analogous relations for the higher
commutators. Indeed, we have identities
\begin{align}
 &[A_{}^{n},\underbrace{[A,\ldots,[A,}_{n-2}B]\ldots]=  \nonumber \\
 &=\dfrac{1}{2_{}^{n}}\sum_{m=1}^{n}\frac{n!(1-(-1)_{}^{m})}{m!(n-m)!}\underbrace{[A_{}^{2},\ldots,[A_{}^{2},}_{n-m}
 \underbrace{[A,\ldots,A,}_{2(m-1)}B]\ldots],\qquad n\geq2,\label{higher1}
\end{align}
so we get the higher linearized equations
\begin{equation}
 \dfrac{\partial_{}^{n-1}B(t)}{\partial t_{1}^{n-2}\partial t_{n}^{}}=\dfrac{1}{2_{}^{n}}
 \sum_{m=1}^{n}\dfrac{n!(1-(-1)_{}^{m})}{m!(n-m)!}\dfrac{\partial_{}^{n+m-2}B(t)}{\partial t_{1}^{2(m-1)}\partial
 t_{2}^{n-m}}, \label{higher2}
\end{equation}
of the KP-hierarchy, where now
\begin{equation}\label{Bt:high}
  B(t_1,t_2,t_n)=e^{t_1A+t_2A^2+t_nA^n}_{}Be^{-t_1A-t_2A^2-t_nA^n}_{}.
\end{equation}
By construction it is obvious that all these flows.

In terms of definitions~(\ref{ad1}) identities~(\ref{commut1}) and~(\ref{higher1}) show that adjoint action
of $A^n$, more exactly $\ad_n(\ad_{1})^{n-2}$, are given in terms of adjoint actions of the lowest powers,
$\ad_1$ and$\ad_2$, that in this sense generate commutative algebra of $\ad_n$'s. Relation~(\ref{commut1})
is lowest in this hierarchy and it corresponds to $n=3$ in~(\ref{higher1}). Indeed, in the case $n=2$ we get
$\ad_2=\ad_2$ and in the case $n=1$ (commuting first Eq.~(\ref{higher1}) with $A$) we get $\ad_1=\ad_1$. Let
us mention also that all identities~(\ref{higher1}) are homogeneous with respect to $A$ and linear with
respect to $B$ by construction.

Relation~(\ref{commut1}) appeared in~\cite{pogr} as equation on the spectral data in study of the inverse scattering transform by means of the
resolvent approach developed in~\cite{first}--\cite{KP-JMP}. Following construction given in~\cite{pogr} it is clear that any integrable equation,
i.e., equation integrable by means of the inverse scattering transform, has in background some algebraic identity describing evolution of scattering
data. In Sec.~2 we present other algebraic identities that generate linearized versions of another integrable equations. In Sec.~3 we introduce a
special realization of arbitrary elements $A$ and $B$ of generic associative algebra that in a natural way enables to derive both, the
\textbf{nonlinear} integrable evolution equation and corresponding Lax pair corresponding to any such identity.

\section{Commutator identities on associative algebras}

Let again $A$ and $B$ be arbitrary elements of an associative algebra and let element $A$ be invertible.
Then again by direct calculation it is easy to check that we have identity
\begin{equation}\label{commut2}
  [A^2,[A^2,[A^{-1},B]]]-[A,[A,[A,[A,[A^{-1},B]]]]]+4[A,[A,[A,B]]]=0,
\end{equation}
Introducing now time dependence, say, as
\begin{equation}\label{Bt2}
  B(t_{1},t_{2},t_{3})=e^{t_{1}A^{-1}+t_{2}A+t_{3}A^2}_{}Be^{-t_{1}A^{-1}-t_{2}A-t_{3}A^2}_{},
\end{equation}
we get that this function of $t$ obeys differential equation
\begin{equation}\label{BLP}
  \dfrac{\partial^{3}B(t)}{\partial t^{2}_{3}\partial t_{1}}-
  \dfrac{\partial^{5}B(t)}{\partial t^{4}_{2}\partial t^{}_{1}}+
  4\dfrac{\partial^{3}B(t)}{\partial t^{3}_{3}}=0,
\end{equation}
that is a linearized version of the Boiti--Leon--Pempinelli (BLP) equation~\cite{BLP} (see also~\cite{GP1}
and~\cite{GP2}).

Identity~(\ref{commut2}) also admits generalization to hierarchy, that in this case is infinite in both
directions. Higher identities are rather complicated and we omit them here mentioning only that identities
for the add powers are simpler, say
\begin{equation}\label{commut2h}
  [A^{3},[A^{-1},B]]-[A,[A,[A,[A^{-1},B]]]]+3[A,[A,B]]=0.
\end{equation}
Then, defining in analogy with~(\ref{Bt2})
\begin{equation}\label{Bt:2}
  B(t_{1},t_{2},t_{3})=e^{t_{1}A^{-1}+t_{2}A+t_{3}A^3}_{}Be^{-t_{1}A^{-1}-t_{2}A-t_{3}A^3}_{},
\end{equation}
we get that this function obeys by~(\ref{commut2h}) differential equation
\begin{equation}\label{BLP2h}
  \dfrac{\partial^{2}B(t)}{\partial t^{}_{3}\partial t_{1}}-
  \dfrac{\partial^{4}B(t)}{\partial t^{3}_{2}\partial t^{}_{1}}+
  3\dfrac{\partial^{2}B(t)}{\partial t^{2}_{2}}=0,
\end{equation}
that is a linearized version of the nonlinear equation suggested in~\cite{GP2}.

In this case commutative algebra of commuting flows given by $\ad_n$ is generated by two adjoint actions:
$\ad_1$ and $\ad_{-1}$.  In order to get identities that give, say, commutator with square in terms of
commutator with the first power, we need at least three elements ($A$, $B_1$ and $B_2$) of an associative
algebra. Let us denote
\begin{align}\label{B:matr}
  &B=\left(\begin{array}{cc}
  0&B_1\\
  B_2&0\end{array}\right),\\
\intertext{and introduce projectors}
  &I_{1}=\left(\begin{array}{cc}
  1&0\\
  0&0\end{array}\right),\qquad
  I_2=\left(\begin{array}{cc}
  0&0\\
  0&1\end{array}\right).\label{I:12}
\end{align}
Then it is easy to check that we have identities
\begin{align}
  &\sigma_{3}[(AI_{1})^{2},B]=[AI_{1},[AI_{1},B]],\label{A+1} \\
  &\sigma_{3}[(AI_{2})^{2},B]=-[AI_{2},[AI_{2},B]],\label{A-1}\\
\intertext{where $\sigma_3=I_1-I_2$ is the standard Pauli matrix. Moreover, if $A$ is invertible, then}
 &[AI_{j},[A^{-1}I_{j},B]]=B,\quad j=1,2.\label{B'}
\end{align}
Now, if we define
\begin{align}\label{Bt:pm}
  &B_{j}(t)=e^{I_{j}(t_1A+t_2A^{2})}_{}Be^{-I_{j}(t_1A+t_2A^{2})}_{},\\
  &B'_{j}(t)=e^{I_{j}(t_1A+t_{-1}A^{-1})}_{}Be^{-I_{j}(t_1A+t_{-1}A^{-1})}_{},\label{B't:pm}
\end{align}
we get differential equations
\begin{align}\label{Bt+:diffe}
  &\sigma_3\partial^{}_{t_2}B^{}_{j}(t)=(-1)^{j+1}\partial^{2}_{t_1}B^{}_{j}(t), \\
  &\partial^{}_{t_1}\partial^{}_{t_{-1}}B'_{j}(t)=B'_{j}(t),\label{Bt:SG}
\end{align}
i.e., linearized Nonlinear Schr\"odinger and Sine-Gordon equations.

It is obvious that relations~(\ref{A+1}) and~(\ref{A-1}) are the lowest members of corresponding
hierarchies, say,
\begin{equation}\label{DS:high}
  \sigma_3^{n}[A^{n}I_{1},B]=\underbrace{[AI_{1},\ldots[AI_{1}}_{n},B]\ldots],\quad n=1,2,\ldots,
\end{equation}
generating hierarchies commuting flows. Relations~(\ref{A+1}) and~(\ref{A-1}) can be used in order to
construct more complicated identities. Let $h$ be a constant diagonal matrix
\begin{equation}\label{h:matr}
  h=\diag\{h_1,h_2\}\equiv h_1I_1+h_2I_2,
\end{equation}
and $B$ as in~(\ref{B:matr}). Then by~(\ref{A+1}) and~(\ref{A-1}) we derive a matrix commutator identities
\begin{align}\label{commut:DS}
  &\sigma_3\bigl[hA^2,B\bigr]=h_1[AI_1,[AI_{1},B]]-
  h_2[AI_{2},[AI_{2},B]],\\
  &[A^{3},B]+3[A,[AI_{1},[AI_{2},B]]]-[A,[A,[A,B]]]=0,\label{commut:VN}\\
\intertext{where}
 &A=AI_1+AI_{2}.\label{A12}
\end{align}
Thus if we define
\begin{align}\label{Bt:DS}
  &B(t_1,t_2,t_3)=e^{t_1AI_{1}+t_2AI_{2}+t_3hA^2}_{}Be^{-t_1AI_{1}-t_2AI_{2}-t_3hA^2}_{},\\
  &B'(t_1,t_2,t_3)=e^{t_1AI_1+t_2AI_2+t_3A^3}_{}Be^{-t_1AI_1-t_2AI_{2}-t_3A^3}_{},\label{Bt:VN}
\end{align}
then thanks to~(\ref{commut:DS})
\begin{align}\label{B:DS}
  &\sigma_3\dfrac{\partial B(t)}{\partial t_3}=h_1\dfrac{\partial^{2}B(t)}{\partial t^{2}_1}-
  h_2\dfrac{\partial^{2}B(t)}{\partial t^{2}_2},\\
\intertext{that is a linearized version of the Davey--Sewartson equations~\cite{DS}. Thanks
to~(\ref{commut:VN}) we derive}
 &\dfrac{\partial B'(t)}{\partial t_3}-\dfrac{\partial^{3} B'(t)}{\partial t_1^{3}}-
 \dfrac{\partial^{3} B'(t)}{\partial t_2^{3}}=0,\label{B':VN}
\end{align}
i.e., a linearized version of the Veselov--Novikov equation~\cite{VN}.

\section{Reconstruction of nonlinear integrable equations by means of commutator identities}
\subsection{Space of operators}\label{resolvent}
We have shown above that commutator identities on associative algebras generate solutions of linearized versions of integrable equations. We also
mentioned that these algebraic identities naturally appear in the framework of the resolvent approach to IST as equations on the scattering data
(see~\cite{pogr} for~(\ref{commut1}) and~\cite{DS1:BPP} for~(\ref{commut:DS})). In this section we demonstrate that these identities enables us to
reconstruct the corresponding nonlinear equations themselves, as well as their Lax pairs. For this aim we need to give some elements of the resolvent
approach (see~\cite{first}--\cite{KP-JMP} for more details). We work in the space of linear integral operators $F(q)$, $G(q)$, etc, with kernels,
correspondingly, $F(x,x';q)$ and $G(x,x';q)$, where $x=(x_1,x_2)$, $x'=(x'_1,x'_2)$, $q=(q_1,q_2)$, and all $x_j$, $x'_j$, and $q_j$ are real
variables. We assume that all these kernels belong to the space of distributions $\Ss'$ with respect to all their six real variables. Thus we can
define the ``shifted'' Fourier transform,
\begin{align}
 F(p;\q) &=\frac{1}{(2\pi)^{2}}\int dx\int dx'\,e_{}^{i(p+{\q}_{\Re})x-i{\q}_{\Re}x'}F(x,x';{\q}_{\Im}),
 \label{2a} \\
 F(x,x';\q_{\Im}) &=\frac{1}{(2\pi)^{2}}\int dp\int d{\q}_{\Re}\,e_{}^{-i(p+{\q}_{\Re})x+i{\q}_{\Re}x'}
 F(p;{\q}),\label{2b}
\end{align}
where $p$ and $\q=\q_{\Re}+i\q_{\Im}$ are respectively real and complex two-dimensional vectors. Vector
$q=\q_{\Im}$ plays role of parameter and it is not touched by composition of such operators
\begin{align}
 &(FG)(x,x';q)=\int dy\,F(x,y;q)G(y,x';q),\label{comp}\\
\intertext{that is defined for pairs of operators for that integral in the r.h.s.\ exists in the sense of
distributions. In the $(p;\q)$-space this composition takes that form}
 &(FG)(p;\q)=\int dp'\,F(p-p';\q+p')G(p';\q).\label{180}
\end{align}

Differential operators form a special subclass of this space of operators. To any given differential
operator $\Lo(x,\partial_{x})$  with the kernel
\begin{equation}
 \Lo(x,x')=\Lo(x,\partial_{x})\delta (x_{1}-x_{1}')\delta(x_{2}-x_{2}')  \label{kern}
\end{equation}
we associate the differential operator $L(q)$ (we call it \textbf{extension} of the differential operator
$\Lo$) with the kernel
\begin{equation}
 L(x,x';q)=e_{}^{-q(x-x')}\Lo(x,x')\equiv \Lo(x,\partial_{x}+q)\delta(x-x'),  \label{1'}
\end{equation}
where $qx=q_{1}x_{1}+q_{2}x_{2}$. Kernel $L(p;\q)$ of the differential operator (as given in~(\ref{2a}))
depends on variables $\q$ polynomially. In particular, let $D_{j}(q)$, $j=1,2$, denote the extension of the
differential operator $i\partial_{x_{j}}$, i.e.,
\begin{align}
 &D_{j}(x,x';q)=i(\partial_{x_{j}}+q_{j})\delta (x-x'),\label{10}\\
\intertext{while in $(p;\q)$-space kernel of this operator equals}
 &D_{j}^{}(p;\q)=\q_{j}^{}\delta (p).  \label{10a}
\end{align}
This property of kernels of differential operators in $(p,\q)$-space suggests introduction of the operation
of the $\partial$-bar differentiation with respect to parameters $\q$:
\begin{equation}
 (\overline{\partial}_{j}^{}F)(p;{\q})=\dfrac{\partial F(p;\q)}
 {\partial\overline{\q}_{j}^{}},\quad j=1,2.  \label{2a12}
\end{equation}
Then characterization property of differential operators is given by condition
\begin{equation}\label{dbL}
  \overline{\partial}_{j}^{}L=0.
\end{equation}
Let us mention that in the case of differential operator with constant coefficients we have by~(\ref{1'}):
\begin{equation}\label{L0}
  L(p;\q)=l(\q)\delta(p),
\end{equation}
where $l(\q)$ is polynomial function of its arguments . Then by~(\ref{180}) we get for commutator with
arbitrary operator $F$ of considered class
\begin{equation}\label{L0F}
  [L,F](p;\q)=(l(p+\q)-l(\q))F(p;\q),
\end{equation}
that is also valid if $l(\q)$ in~(\ref{L0}) is meromorphic function.

We skip here additional conditions that enables introduction and unique definition of inverse operators and
for future use mention only that operator inverse to $D_j$ will be defined as (cf.~(\ref{10a}))
\begin{equation}\label{Dj-1}
  D_{j}^{-1}(p;\q)=\dfrac{\delta (p)}{\q_{j}^{}},
\end{equation}
while its kernel in the $x$-space is given by~(\ref{2b}).

\subsection{(2+1)-dimensional integrable equations}\label{2+1}
Now we introduce a specific realization of  elements $A$ and $B$ of associative algebra that enables
derivation of the Lax pairs and nonlinear integrable equations. In this section we consider
identities~(\ref{commut1}), (\ref{higher1}), (\ref{commut2}), (\ref{commut2h}), (\ref{commut:DS}) (in the
case $h_1h_2\neq0$) and~(\ref{commut:VN}), i.e., those identities that are generated by two adjoint actions
of powers of $A$ (or its matrix functions as in~(\ref{commut:DS}) and~(\ref{commut:VN})). Correspondingly,
we realize $A$ and $B$ (or $B_1$, $B_2$) as integral operators $A(q)$ and $B(q)$ in two-dimensional space in
the sense of definition in Sec.~\ref{resolvent}. Thus dimension of variables $x$ and $x'$ of kernels is
chosen to be equal to the number of generators of the identities under consideration. These operators are
labeled by two dimensional real vector $q$ with kernels $A(x,x';q)$ and $B(x,x';q)$. We impose condition
that evolution with respect to any time $t_m$ preserves this property, then switching on times by means of
one of relations~(\ref{Bt1}), (\ref{Bt:high}), (\ref{Bt2}), (\ref{BLP2h}), (\ref{Bt:DS}), or~(\ref{Bt:VN})
gives operators $B(t,q)$ with kernels $B(x,x';t,q)$ belonging to the same space of operators. Taking into
account that time variables corresponding to two generators of these identities were denoted as $t_1$ and
$t_2$, we impose the following two conditions and show that they determine operators $A(q)$ and $B(q)$.

\textbf{Condition 1.} The time evolution of $B(q)$ with respect to variables $t_1$ and $t_2$ is just shift
of $x$-variables of the kernel, i.e.,
\begin{equation}\label{cond:tx}
  B(x,x';t_1,t_2,q) =B(x_1+t_1,x_2+t_2,x'_1+t_1,x'_2+t_2;q),
\end{equation}
where dependence on other $t_m$-variables was omitted.

In terms of $(p,\q)$-kernels defined by~(\ref{2a}) this means that
\begin{align}
 &B(p;t,\q)=e^{-it_1p_1-it_2p_2}B(p;\q).\label{cond:tp}\\
\intertext{so that in the differential form we have}
 \label{Bn1}
  &\partial_{t_{j}}B(t,x,x';q)=(\partial_{x_j}+\partial_{x'_j})B(t,x'x';q),\quad j=1,2,\\
  &\partial_{t_{j}}B(t,p;\q)=-ip_{j}B(t,p;\q),\label{Bn2}
\intertext{that thanks to~(\ref{10}) or~(\ref{10a}) and~(\ref{L0F}) can be written as}
 &i\partial_{t_{j}}B(t,q)=[D_j(q),B(t,q)].\label{Bn3}
\end{align}
In order to motivate this condition we mention that in~\cite{pogr} in the study in terms of IST of the
algebraic schemes developed in~\cite{DJKM}--\cite{Orlov} relation~(\ref{cond:tx}) was derived as evolution
of operator scattering data.

\textbf{Condition 2.} If dependence of the kernel $A(p;\q)$ on $\q$-variables reduces to only one linear
combination of $\q_1$ and $\q_2$ with constant coefficients, than the same is $\q$-dependence of the kernel
$B(p;\q)$.

\subsubsection{KPII-equation}
We illustrate procedure of reconstruction of the Lax pair and nonlinear equation using
identity~(\ref{commut1}) as example. By~(\ref{Bt1}) we get that
\begin{equation}\label{Bt1'}
  \partial B(t)/\partial t_{1}=[A,B(t)],\qquad\partial B(t)/\partial t_{2}=[A^2,B(t)].
\end{equation}
Thus, the first equality here and~(\ref{Bn3}) show that it is natural to choose
\begin{equation}\label{A:KPII}
  A(p;\q)=-i\q_1\delta(p),\quad\text{or}\quad A=-iD_1.
\end{equation}
Now the second equality in~(\ref{Bt1'}) shows that thanks to~(\ref{L0F}) and~(\ref{Bn2}) the kernel of
operator $B(q)$ in $(p;\q)$-space obeys
\begin{equation}\label{KPII:B}
  [ip_2+(p_1+\q_1)^2-\q^2_1]B(p;\q)=0.
\end{equation}
By means of~(\ref{L0F}) this can be written as equality
\begin{equation}\label{L0B}
  [L_0(q),B(q)]=0,
\end{equation}
where the kernel of operator $L_0$ equals
\begin{align}\label{KPII:L0p}
 &L_{0}(p;\q)=(i\q_2+\q^2_1)\delta(p),\\
\intertext{so that taking into account definitions~(\ref{10a}) we can write this operator as}
 &L_{0}=iD_2+D^2_1.\label{KPII:L0}\\
\intertext{By~(\ref{kern})--(\ref{10}) this operator is extension of operator}
 &\Lo_0=-\partial^{}_{x_2}-\partial^{2}_{x_1},\label{KPII:L0'}\\
\end{align}
that is the differential part (part corresponding to zero potential) of the Lax operator associated to
the KPII equation (see,~\cite{ZS,Dr}):
\begin{equation}\label{KPII:L}
  \Lo=-\partial^{}_{x_2}-\partial^{2}_{x_1}-u(x).
\end{equation}

Condition~2 means that kernel $B(p;\q)$ is independent of $\q_2$, so that by~(\ref{KPII:B}) we have that it
has the form
\begin{equation}\label{Bb}
  B(p;t,\q)=\delta(ip_2+p_1(p_1+2\q_1))b(p,t),
\end{equation}
(we skip here consideration of more complicated solutions of~(\ref{KPII:B}) that are given by
$\partial/\partial\overline\q_1$-derivatives of~(\ref{Bb})), where we used $\delta$-function of complex
argument, $\delta(z)=\delta(z_{\Re})\delta(z_{\Im})$. Here $b(p,t)$ is a function of $p$ and $t$, that
equals
\begin{equation}\label{b(t)}
  b(p,t_1,t_2,t_3)=\exp\left(-ip_1t_1-ip_2t_2+it_3\dfrac{p^4_1-3p^2_2}{4p_1}\right)b(p),
\end{equation}
when all $t_1$, $t_2$, and $t_3$ are switched on by~(\ref{Bt2}) and where $b(p)$ is independent of $t$ (and
$\q$).

Now we introduce operator $\nu$ with kernel $\nu(p;\q)$ depending on the same linear combination of $\q_1$
and $\q_2$ that was mentioned in Condition~2 by means of $\overline\partial$-problem with respect to this
linear combination and given in terms of the operator $B(q)$. In our case thanks to~(\ref{A:KPII})
and~(\ref{Bb}) this means that $\nu(p;\q)=\nu(p;\q_1)$ and
\begin{equation}
 \overline{\partial}_{1}^{}\nu=\nu B,\label{26}
\end{equation}
(see~(\ref{2a12})). In order to determine operator $\nu$ uniquely we normalize it to be unity operator at
point of singularity of the kernel $L_0(p;\q)$ with respect to $\q_1$. Here by~(\ref{A:KPII}) this means:
\begin{equation}
 \nu(p;\q_1)=\delta(p)+O(\q^{-1}_1),\quad \q_1\to\infty.\label{2520}
\end{equation}
Then thanks to~(\ref{dbL}) we get from~(\ref{L0B}) that $[L_0,\nu]$ obeys
\begin{equation}
 \overline{\partial}_{1}^{}[L_0,\nu]=[L_0,\nu]B,\label{L0nu}
\end{equation}
i.e., the same Eq.~(\ref{26}). By~(\ref{L0F}) and~(\ref{2520}) kernel $[L_0,\nu](p;\q_1)$ has some constant
(with respect to $\q_1$) asymptotic behavior at infinity, that we denote $-u(p,t)$. Thus
\begin{equation}
 L\nu=\nu L_{0},\label{23}
\end{equation}
where $L$ is extension (in the sense of~(\ref{1'})) of differential operator~(\ref{KPII:L}). Eq.~(\ref{23})
means that $\nu$ is dressing (transformation) operator. In~\cite{KPIIreg} it was shown that under
inserting~(\ref{Bb}) into~(\ref{26}) we obtain the standard (see~\cite{dbar}) equation of the inverse
problem for the Jost solution, and that the second operator of the Lax pair as well as differential equation
on $u(x,t)$ (the Fourier transform of $u(p,t)$) follow from~(\ref{b(t)}) and~(\ref{26}). We do not reproduce
here these details.

\subsubsection{KPI-equation}

Above we derived KPII equation under assumption that all times in~(\ref{Bt1}) are real. Performing
substitution $t_{j}\to-it_{j}$ in~(\ref{Bt1}) and~(\ref{higher1}) we get instead of~(\ref{Bt1'})
\begin{equation}\label{Bt1''}
  i\partial B(t)/\partial t_{1}=[A,B(t)],\qquad i\partial B(t)/\partial t_{2}=[A^2,B(t)],
\end{equation}
and so on. Thus now in the same way as~(\ref{A:KPII}) we derive that
\begin{equation}\label{A:KPI}
  A=D_1,
\end{equation}
and thanks to~(\ref{L0F}), (\ref{Bn2}) and the the second equality in~(\ref{Bt1''}) in this case the kernel
of operator $B(q)$ in $(p;\q)$-space obeys
\begin{equation}\label{KPI:B}
  [p_2-(p_1+\q_1)^2+\q^2_1]B(p;\q)=0.
\end{equation}
This means that operator $B(q)$ obeys the same commutator identity~(\ref{L0B}), while now
\begin{equation}\label{KPI:L0}
  L_{0}=D^{}_2-D^2_1.
\end{equation}
In terms of Sec.~\ref{resolvent} this is an extension of differential operator
\begin{equation}\label{KPI:L0'}
  \Lo_0=i\partial^{}_{x_2}+\partial^{2}_{x_1},
\end{equation}
i.e., of the differential part of the Lax operator associated to the equation KPI (see,~\cite{ZS,Dr}):
\begin{equation}\label{KPI:L}
  \Lo=i\partial^{}_{x_2}+\partial^{2}_{x_1}-u(x).
\end{equation}
Now, thanks to Condition~2 and in analogy to~(\ref{Bb}) and~(\ref{b(t)}) we get
\begin{align}\label{Bb1}
  &B(t,p;\q_1)=\delta(p_2-p_1(p_1+2\q_1))b(t,p),\\
\intertext{where}
 \label{b1(t)}
  &b(t_1,t_2,t_3,p)=\exp\left(-ip_1t_1-ip_2t_2-it_3\dfrac{p^4_1+3p^2_2}{4p_1}\right)b(p).
\end{align}

Again we can define transformation operator $\nu$ by means of~(\ref{26}) and~(\ref{2520}). But in this case
the argument of $\delta$-function in the r.h.s.\ of~(\ref{Bb1}) is proportional to $\q_{1\Im}=q_1$, so that
instead of $\overline\partial$-problem~(\ref{26}) we get nonlocal Riemann--Hilbert problem---well known
difference of the inverse problems for KPII and KPI equations, cf~\cite{dbar} with~\cite{ZM}
and~\cite{Manakov}. Again here $L_0(p;\q)$ behaves polynomially at infinity, so we impose the same
normalization~(\ref{2520}). Next, thanks to~(\ref{L0B}) we prove relation~(\ref{23}) where now $L$ is
extension in the sense of~(\ref{1'}) of the differential operator~(\ref{KPI:L}). Referring now
to~\cite{KPIreg} one can derive the second equation of the Lax pair and prove $u(x,t)$ obeys KPI equation
when three times, $t_1$, $t_2$, and $t_3$ are switched on. In the same way we can use
identity~(\ref{higher1}) to derive higher equations of KP hierarchy, see~\cite{pogr}. By construction it is
clear that all of them have the same associated linear operator $\Lo$.

\subsubsection{BLP equation}

Let now $B(t)$ to be defined by~(\ref{Bt2}), so that by~(\ref{commut2}) it obeys Eq.~(\ref{BLP}). Then
instead of~(\ref{Bt1'}) we get
\begin{equation}\label{Bt:BLP}
  \partial B(t)/\partial t_{1}=[A,B(t)],\qquad\partial B(t)/\partial t_{2}=[A^{-1},B(t)].
\end{equation}
Using Condition~1, or one of its differential forms~(\ref{Bn1})--(\ref{Bn3}) we obtain that operator $A$ can
be chosen like in~(\ref{A:KPII}). Then in analogy with derivation of Eq.~(\ref{KPII:B}) we get by the second
equality in~(\ref{Bt:BLP}), (\ref{L0F}), and~(\ref{Dj-1}) for $j=2$ that here kernel of operator $B$ obeys
\begin{equation}\label{BLP:B}
  \left[p_2+\dfrac{1}{p_1+\q_1}-\dfrac{1}{\q_1}\right]B(p;\q)=0,
\end{equation}
Thus again $B(q)$ obeys Eq.~(\ref{L0B}), where by~(\ref{L0F}) operator $L_0$ equals
\begin{equation}\label{BLP:L0}
  L_{0}=D^{}_2+D^{-1}_1.
\end{equation}
Condition~2 gives then again that $B(p;\q)=B(p;\q_1)$ and relation~(\ref{BLP:B}) means that kernel $B(p;\q)$
has the form (cf.~(\ref{Bb}))
\begin{align}\label{BLP:Bb}
  &B(t,p;\q)=\delta\biggl(p_2+\dfrac{1}{p_1+\q_1}-\dfrac{1}{\q_1}\biggr)b(t,p),
\intertext{where $b(t,p)$ is a function of $p$ and $t$ equal}
 \label{BLP:b(t)}
  &b(t_1,t_2,t_3,p)=\exp\left(-ip_1t_1-ip_2t_2-it_3p^2_1\sqrt{1-\dfrac{4}{p_1p_2}}\,\right)b(p),
\end{align}
when all $t_1$, $t_2$, and $t_3$ are switched on by means of~(\ref{Bt2}).

Kernel of operator $L_0$ equals $L_0(p;\q)=(\q_2+1/\q_1)\delta(p)$, so in contrast to the KP case it is not
entire but only meromorphic function of $\q_1$. Correspondingly, dressing operator $\nu$ with kernel
$\nu(p;\q_1)$ is defined by the same $\overline\partial$-problem~(\ref{26}) but now $L_0(p;\q)$ is singular
at $\q_1=0$, so normalization condition is given as
\begin{equation}
 \nu(p;\q_{1})\Bigr|_{\q_1=0}=\delta(p),\label{25}
\end{equation}
and we look for solutions bounded at $\q_1\to\infty$. As a consequence, $\overline\partial$-equation on
commutator $[L_0,\nu]$ has additional inhomogeneous (cf.~(\ref{L0nu})). Nevertheless, after simple
calculations we prove that Eq.~(\ref{23}) is valid also in this case, where dressed operator $L$ equals
\begin{equation}\label{BLP:L}
  L=D^{}_2+D^{-1}_1+D^{-1}_1\alpha+\beta,
\end{equation}
where $\alpha$ and $\beta$ are multiplication operators, i.e., in the $x$-space they have kernels
$\alpha(x,x';q)=\alpha(x)\delta(x-x')$ and the same for $\beta$. Operator~(\ref{BLP:L}) is not a
differential one, in order to get differential operator we have to consider product $D_1L$ that thanks
to~(\ref{1'}) is extension of operator
\begin{equation}\label{L-blp}
  \widetilde\Lo=\partial_{x_1}\partial_{x_2}+1+\alpha_{x}+\partial_{x_1}\cdot\beta(x),
\end{equation}
i.e., auxiliary linear operator of the BLP equation, see~\cite{BLP}. In~\cite{GP1} it was mentioned that
study of direct and inverse problems for this operator, in fact, reduces to study of these problems for
operator~(\ref{BLP:L}). In~\cite{GP2} it was proved that time evolution (with respect to $t_3$) of spectral
data $B$ leads by~(\ref{BLP:b(t)}) to nonlinear differential BLP-equation on $\alpha$ and $\beta$.

Exponent in Eq.~(\ref{BLP:b(t)}) is not pure imaginary. Correspondingly, time evolution for BLP equation is
unstable, see~\cite{GP2}. It was also proved there that if dependence on $t_3$ is given by~(\ref{Bt:2}) then
resulting nonlinear differential equation on $\alpha$ and $\beta$ has no this defect. Here we omit further
details and mention only that dependence on $t_1$ and $t_2$ in~(\ref{Bt2}) and~(\ref{Bt:2}) coincides, so in
the case of identity~(\ref{BLP2h}) we derive by means of the above procedure the same
$L$-operator~(\ref{BLP:L}).

\subsubsection{DS and VN equations}
Now we consider briefly consequences of identities~(\ref{commut:DS}) and~(\ref{commut:VN}), where $B$ is a
matrix given in~(\ref{B:matr}) and matrices $I_j$ are defined in~(\ref{I:12}). Let $B_1$ and $B_2$ be
operators in the sense of Sec.~\ref{resolvent} with kernels $B_j(x,x';q)$ in $x$-space (correspondingly
$B_{j}(p;\q)$ in $(p,\q)$-space). Under time evolutions~(\ref{Bt:DS}) and~(\ref{Bt:VN}) we get,
correspondingly operators with kernels $B(t,x,x';q)$ and $B'(t,x,x';q)$. Let us consider the first equality.
We impose on the matrix operator $B(t,q)$ Condition~1 and then by~(\ref{Bn2}) and~(\ref{I:12}) we derive
\begin{align}\label{AB:BA1}
  &i(AB_1)(p;\q)=p_1B_1(p;\q),&&-i(B_2A)(p;\q)=p_1B_2(p;\q), \\
  &-i(B_1A)(p;\q)=p_2B_1(p;\q),&&i(AB_2)(p;\q)=p_2B_2(p;\q).\label{AB:BA2}
\end{align}
Summing equations in columns we get that $i[A,B_j](p;\q)=(p_1+p_2)B_j(p;\q)$, $j=1,2$, so that
by~(\ref{L0F}) here we have to put
\begin{align}\label{DS:A1}
 &A(p;\q)=-i(\q_1+\q_2)\delta(p),\\
\intertext{that can be written in the operator form thanks to~(\ref{10a}) as}
  &A=-i(D_1+D_2).
\end{align}
Now we are left to obey, say, equations in~(\ref{AB:BA1}) that gives conditions
$(\q_1+\q_2+p_2)B_{1}(p;\q)=0$ and $(\q_1+\q_2+p_1)B_{2}(p;\q)=0$. To obey these equalities it is enough to
choose the kernel of operator $B$ in the form
\begin{equation}\label{DS:B}
  B(p;t,\q)=\left(\begin{array}{cc}
  0&\delta(\q_1+\q_2+p_2)b_1(p,t)\\
  \delta(\q_1+\q_2+p_1)b_2(p,t)&0\end{array}\right),
\end{equation}
where we used that by Condition~2 and~(\ref{DS:A1}) the kernel $B(t,p;\q)$ is independent of the difference
$\q_1-\q_2$. Time dependence of operator $B$ in the case when all three times are switched on
by~(\ref{Bt:DS}) is given by means of explicit relation
\begin{align}\label{DS:B1}
  B(p;t,\q)&=e^{-it_1p_1-it_2p_2-t_3\sigma_3(h_1p^2_1-h_2p^2_2)}\times\nonumber\\
  &\times\left(\begin{array}{cc}
  0&\delta(\q_1+\q_2+p_2)b_1(p)\\
  \delta(\q_1+\q_2+p_1)b_2(p)&0\end{array}\right),
\end{align}
so that in order to avoid instability we have to change $t_3$ to $it_3$.

Now we have to construct nontrivial operator $L_0$ with constant coefficients that obeys
equation~(\ref{L0B}). Taking into account $\delta$-functions in~(\ref{DS:B}) and off-diagonal structure of
operator $B$, we choose $L_0(p;\q)=\diag\{P_1(\q),P_2(\q)\}\delta(p)$, where $P_j(\q)$ are polynomials.
Under condition that these polynomials are nontrivial and of the lowest possible degree, we easily find that
\begin{equation}\label{DS:L0}
  L_0=\left(\begin{array}{cc}
  D_2&0\\
  0&-D_1\end{array}\right).
\end{equation}
Thanks to our definitions~(\ref{kern})--(\ref{10a}) this is extension of operator
\begin{equation}\label{DS:L01}
  \Lo_0=\left(\begin{array}{cc}
  \partial_{x_2}&0\\
  0&-\partial_{x_1}\end{array}\right),
\end{equation}
that is the differential part of the two-dimensional Zakharov--Shabat problem, known to be an auxiliary
linear problem for the DS equation~\cite{FA}.

Construction of dressing operator $\nu$ with kernel $\nu(p;\q_1+\q_2)$ can be performed by means of
$\overline\partial$-problem~(\ref{26}) normalized to be unity matrix operator (cf.~(\ref{2520})) at point of
singularity of the kernel $L_0(p;\q)$, i.e., at $\q_1+\q_2\to\infty$. Then we prove that Eq.~(\ref{L0nu}) is
valid thanks to~(\ref{L0B}) and~(\ref{DS:L0}). Then, taking into account the asymptotic behavior of the
kernel $[L_0,B](p;\q)$ at $\q$-infinity we derive~(\ref{23}), where now $L$ is extension of the linear
operator of the Zakharov--Shabat problem, see~\cite{DS1:BPP} for details. Consideration of
identity~(\ref{commut:VN}) goes along the same lines, as dependencies on $t_1$ and $t_2$ in~(\ref{Bt:DS})
and~(\ref{Bt:VN}) coincide.

\subsection{(1+1)-dimensional equations}
Above we have seen that without lost of generality operator $A$ can be always chosen to be a linear
combination with constant (matrix) coefficients of operators $D_1$ and $D_2$ (see~(\ref{10}), (\ref{10a})).
In order to get integrable equations in $(1+1)$ dimensions we can perform reduction, imposing condition that
operator $B$ commutes with additional linear combination of $D_1$ and $D_2$. Then as follows
from~(\ref{L0F}) function $b(p)$ in~(\ref{b(t)}), (\ref{b1(t)}), and~(\ref{BLP:b(t)}) is proportional to
$\delta(p_2)$ and functions $b_j(p)$ in~(\ref{DS:B}) are proportional to $\delta(p_1+p_2)$. It is easy to
check that in this way we indeed arrive to corresponding (1+1)-dimensional nonlinear evolution equations. On
the other side we can directly apply the procedure described above to identities~(\ref{A+1})--(\ref{B'}) and
corresponding equations~(\ref{Bt+:diffe}) and~(\ref{Bt:SG}). Let us consider as example Eq.~(\ref{A+1}).
Here we have only one generator of commutator algebra, i.e., commutator with $AI_1$. Correspondingly, we
realize operator $B$ as integral operator in one dimensional space: operator with kernel $B(x_1,x'_1;q)$.
Using obvious corresponding one dimensional modification of techniques given in Sec.~\ref{resolvent} we
impose here instead of Condition~1 (see~(\ref{cond:tx})) condition
\begin{align}\label{cond:tx1}
  &B(x_1,x'_1;t_1,q) =B(x_1+t_1,x'_1+t_1,;q_1),\\
\intertext{so that by~(\ref{Bn2}) for $j=1$:}
 &\partial_{t_{1}}B(p_1;t_1,\q_1)=-ip_{1}B(p_1;t_1,\q_1).
\end{align}
On the other side by~(\ref{Bt:pm}) $\partial_{t_{1}}B(t)=[AI_1,B(t)]$. We choose $A$ to be a differential
operator with constant coefficients, so thanks to~(\ref{L0}) we have for its kernel representation
$A(p_1,\q_1)=a(\q_1)\delta(p_1)$ where $a(\q_1)$ is polynomial. Thanks to~(\ref{B:matr}), (\ref{I:12}), and
property~(\ref{L0F}) we get
\begin{equation}\label{1dim}
   (a(p_1+\q_1)+ip_1)B_1(p_1,\q_1)=0,\qquad (a(\q_1)-ip_1)B_2(p_1,\q_1)=0.
\end{equation}
In this case we have no variables $p_2$ and $\q_2$, so zeroes of expressions in parenthesis must occur for
the same values of $\q_1$ and $p_1$. This condition defines $a(\q_1)$ to be a polynomial of the first order
equal (up to a constant shift) to $-2i\q_1$, i.e.,
\begin{align}\label{1dim:A}
  & A(p_1;\q_1)=-2i\q_1\delta(p_1) \\
\intertext{or by~(\ref{10a})}
  & A=-2iD_1.\label{1dim:A1}\\
\intertext{Then equalities~(\ref{1dim}) reduce to $(p_1+2\q_1)B_j(p_1,\q_1)=0$, so that by~(\ref{B:matr}) we
have for the kernel of operator $B$ representation}
 \label{1dim:B}
  &B(t,p_1;\q_1)=\left(\begin{array}{cc}
  0&b_1(t,p_1)\\
  b_2(t,p_1)&0\end{array}\right)\delta(p_1+2\q_1).
\end{align}
As above it is easy to prove that operator $L_0$ that obeys relation~(\ref{L0B}) is given by
\begin{equation}\label{1dim:L0}
  L_0=\sigma_3 D_1,
\end{equation}
so that we arrive to the differential part of the one dimensional Zakharov--Shabat problem. Construction of
dressing operator $\nu$ and potential $u$ can be performed along the same lines as above by means of the
$\overline\partial$-bar problem~(\ref{26}) that here like in the cases KPI and DS above reduces to the
Riemann--Hilbert problem thanks to specific argument of the $\delta$-function in~(\ref{1dim:B}).

\section{Concluding remarks}

In this article we demonstrated that existence of a commutator identities leads to integrable nonlinear
evolution equation. suggested scheme of reconstruction of such equation together with its Lax operators is
generic and model independent, while examples considered here were corresponding to the known integrable
equations only. In this sense it is interesting and opened problem to give description of all commutator
identities on associative algebras. Another opened problem is existence of commutator identities with three
generators. Following the lines of the suggested here procedure it is reasonable to expect that such
identity can lead to (3+1)-dimensional integrable equation. Commutator identities described here are of
generic form, while realization of operators $A$ and $B$ that enabled us to get nonlinear evolution
equations was rather specific. It is worth to assume that the same identities are related with some other
integrable systems under some other realization of elements of associative algebra.


\begin{thebibliography}{99}

\bibitem{pogr} A.~K.~Pogrebkov, "On time evolutions associated with the nonstationary
Schro"dinger equation" in L.~D.~Faddeev's Seminar on Mathematical Physics, Ed.\
M.~Semenov-Tian-Shansky, {\sl Amer.\ Math.\ Soc.\ Transl.\/} {\bf (2) 201} pp.
239--255 (2000)
\bibitem{KP} B.~B.~Kadomtsev and V.~I.~Petviashvili, {\sl Soviet Phys. Dokl.\/} {\bf 192} (1970), 539
\bibitem{first}  M.~Boiti, F.~Pempinelli, A.~K.~Pogrebkov, and M.~C.~Polivanov, {\sl Theor. Math. Phys.\/} {\bf 93}
1200 (1992).
\bibitem{total}  M.~Boiti, F.~Pempinelli, A.~K.~Pogrebkov, and M.~C.~Polivanov, {\sl Inverse Problems\/} {\bf 8} 331 (1992).
\bibitem{KPIreg}  M.~Boiti, F.~Pempinelli, and A.~Pogrebkov, {\sl Journ. Math. Phys. \/} {\bf 35} 4683 (1994).
\bibitem{KPIIreg} M.~Boiti, F.~Pempinelli, A.~K.~Pogrebkov and B.~Prinari, {\sl Inverse Problems\/} {\bf 17} 937 (2001)
\bibitem{KP-JMP} M.~Boiti, F.~Pempinelli, A.~K.~Pogrebkov and B.~Prinari, {\sl Journ. Math. Phys.\/} {\bf 44} 3309 (2003)
\bibitem{BLP} M.~Boiti, J.~Leon, F.~Pempinelli, {\sl Inverse Problems} {\bf 3} (1987) 37--49.
\bibitem{GP1} T.~I.~Garagash, A.~K.~Pogrebkov, {\sl Theor. Math.Phys.\/} {\bf 102} (1995) 117--132.
\bibitem{GP2} T.~I.~Garagash, A.~K.~Pogrebkov, {\sl Theor. Math.Phys.\/} {\bf 109} (1997) 1369--1379.
\bibitem{DS} A.~Davey and K.~Stewartson, {\sl Proc. Royl Soc.\/} {\bf A338} (1974) 101.
\bibitem{VN} A.~P.~Veselov and S.~P.~Novikov, {\sl Soviet Math. Dokl.\/} {\bf 30} (1984), 588–-591.
\bibitem{DS1:BPP} M.~Boiti, F.~Pempinelli, and A.~K.~Pogrebkov, ``DSI revisited,'' in preparation.
\bibitem{DJKM}  E.~Date, M.~Jimbo, M.~Kashiwara, and T.~Miwa, ``Transformation
Groups for Soliton Equations,'' in: {\sl Nonlinear Integrable Systems\/{\rm:} Classical Theory and Quantum Theory,} ed. M.~Jimbo and T.~Miwa,
(Singapore: World Scientific, 1983).
\bibitem{OrlovS}  A. Yu. Orlov and E. I. Schulman, {\sl Lett. Math. Phys.} {\bf 12} (1986): 171.
\bibitem{Orlov}  A.~Yu.~Orlov, ``Vertex Operator, $\bar{\partial}$-Problem, Symmetries, Variational Identities, and Hamiltonian Formalism for
(2+1)-Dimensional Integrable Equations,'' in: {\sl Proc. Intl. Workshop ``Plasma Theory and Nonlinear and Turbulent Processes in Physics''}, Vol. 1,
ed.~V.G.Bar'yakhtar, V.~M.~Chernousenko, N.~S.~Erokhin, A.~G.~Sitenko, and V.~E.~Zakharov (Singapore: World Scientific, 1988), p. 116.
\bibitem{ZS} V.~E.~Zakharov and A.~B.~Shabat, \textsl{Funct. Anal. Appl.\/} \textbf{8} (1974) 226
\bibitem{Dr} V.~S.~Dryuma, \textsl{Sov. Phys. J. Exp. Theor. Phys. Lett.\/} \textbf{19} (1974) 381
\bibitem{dbar} M.~J.~Ablowitz, D.~Bar~Yacoov and A.~S.~Fokas, \textsl{Stud. Appl. Math.\/} \textbf{69} (1983) 135
\bibitem{ZM} V.~E.~Zakharov, S.~V.~Manakov, \textsl{Sov. Scien. Reviews\/} \textbf{1} (1979) 133
\bibitem{Manakov} S.~V.~Manakov, \textsl{Physica\/} \textbf{D3} (1981) 420
\bibitem{FA} A.~S.~Fokas and M.~J.~Ablowitz, \textsl{Phys. Rev. Lett.\/} \textbf{51} (1983) 7
\end{thebibliography}
\end{document}